\documentstyle[psfig,11pt]{article}
\begin{document}
\vspace{1.0cm}
{\Large \bf STUDYING THE HIGH-ENERGY GAMMA-RAY SKY WITH GLAST}

\vspace{1.0cm}

T. Kamae$^1$, T. Ohsugi$^2$, D.J. Thompson$^3$ and K. Watanabe$^4$ 

\vspace{1.0cm}
$^1${\it Dept. of Physics, Univ. of Tokyo, Bunkyo-ku, Tokyo, Japan \\
kamae@phys.s.u-tokyo.ac.jp}\\
$^1${\it Dept. of Physics, Hiroshima Univ., Higashi-Hiroshima, Hiroshima, Japan \\
ohsugi@hirax3.hepl.hiroshima-u.ac.jp}\\
$^3${\it LHEA, NASA Goddard Space Flight Center,Greenbelt, Maryland  USA \\
djt@egret.gsfc.nasa.gov}\\
$^4${\it USRA/LHEA, NASA Goddard Space Flight Center,Greenbelt, Maryland  USA \\
watanabe@grossc.gsfc.nasa.gov}\\

{\bf on behalf of the GLAST Collaboration}

(A joint paper of E1.1-0062 and E1.1-0063)

\vspace{0.5cm}

\section*{ABSTRACT}
Building on the success of the Energetic Gamma Ray Experiment Telescope (EGRET)
on the Compton Gamma Ray Observatory, the Gamma-ray Large Area Space Telescope 
(GLAST) will make a major step in the study of such subjects as blazars, 
gamma-ray bursts, the search for dark matter, supernova remnants, pulsars, 
diffuse radiation, and unidentified high-energy sources. The instrument will be
built on new and mature detector technologies such as silicon strip detectors, 
low-power low-noise LSI, and a multilevel data acquisition system. GLAST is in 
the research and development phase, and one full tower (of 25 total) is now 
being built in collaborating institutes. The prototype tower will be tested 
thoroughly at SLAC in the fall of 1999.

\section{INTRODUCTION}
As the highest-energy photons, gamma rays have an inherent interest to 
astrophysicists and particle physicists studying high-energy, nonthermal 
processes.  Gamma-ray telescopes complement those at other wavelengths, 
especially radio, optical, and X-ray, providing the broad, mutiwavelength 
coverage that has become such a powerful aspect of modern astrophysics.  EGRET,
the high-energy telescope on the Compton Gamma Ray Observatory, has led the way
in such an effort, contributing to broad-band studies of blazars, gamma-ray 
bursts, pulsars, solar flares, and diffuse radiation. Now development is 
underway for the next significant advance in high-energy gamma-ray astrophysics,
GLAST, which will have $\sim$30 times the sensitivity of EGRET at 100 MeV and 
more at higher energies, including the largely-unexplored 30--300 GeV band.  
The following sections describe the science goals, instrument technologies, and
international collaboration for GLAST.

Some key scientific parameters for GLAST are shown in Table 1 
(Bloom, 1996; Michelson, 1996). 

\section{SCIENTIFIC GOALS FOR GLAST}

\subsection{\underline{Blazars}}
Blazars are thought to be active galactic nuclei consisting of accretion-fed 
supermassive black holes with jets of relativistic particles directed nearly 
toward our line of sight.  The formation, collimation, and particle 
acceleration in these powerful jets remain important open questions. Many 
blazars are seen as bright, highly-variable gamma-ray sources, with the 
high-energy gamma rays often dominating the luminosity (Hartman et al. 1997).  
For this reason, the gamma rays provide a valuable probe of the physics under 
these extreme conditions, especially when studied as part of multiwavelength 
campaigns (e.g. Shrader and Wehrle, 1997). With its wide field of view and high 
sensitivity, GLAST will enable blazar studies with far better resolution and 
time coverage than was possible with EGRET. GLAST should detect thousands of
blazars.

Blazars are often very distant objects, and the extension of the gamma-ray 
spectrum into the multi-GeV range opens the possibility of using blazars as 
cosmological probes.  In the energy range beyond that observed by EGRET but 
accessible to GLAST, the blazar spectra should be cut off by absorption 
effects of the extragalactic background light produced by galaxies during the 
era of star formation.  A statistical sample of high-energy blazar spectra at 
different redshifts may provide unique information on the early universe 
(Macminn and Primack, 1996). 

\begin{table}
\centering
\caption{\label{1} SOME GLAST CHARACTERISTICS}
\smallskip
\begin{tabular}{ll}
\hline
            &  \\
Energy Range  & 20 MeV -- 300 GeV\\
& \\
Energy Resolution  & 10\% ($>$100 MeV)\\
& \\
Effective Area &	   $>$8000 cm$^2$ ($>$1 GeV) \\
& \\
Field of view 	& $>$2.5 steradians \\
& \\
Source location determination	& 30 arc seconds to 5 arc minutes\\
& \\
Source sensitivity (1 yr.)	& 3 x 10$^{-9}$ ph (E$>$100 MeV)/cm$^2$s
\\
& \\
Mission life	& $>$ 2 years\\
\hline
\end{tabular}
\end{table}

\subsection{\underline{Gamma-Ray Bursts}}
The recent breakthrough associating gamma-ray bursts with distant galaxies 
(e.g. Djorgovski et al. 1997) has changed the focus of gamma-ray burst 
research from the question of where they are to the questions of what they are 
and how they work.  The power source and emission mechanisms for gamma-ray 
bursts remain mysteries.  The high-energy gamma radiation seen from some 
bursts by EGRET (Dingus et al. 1995) indicates that GLAST will provide 
important information about these questions.  With its large field of view, 
GLAST can expect to detect over 100 bursts per year at GeV energies compared to
about one per year for EGRET, allowing studies of the high-energy component of 
the burst spectra. 

\subsection{\underline{Search for Dark Matter}}
One of the leading candidates for the dark matter now thought to dominate the 
universe is a stable, weakly-interacting massive particle (WIMP).  One 
candidate in supersymmetric extensions of the standard model in particle 
physics is the neutralino, which might annihilate into gamma rays in the 
30-300 GeV range covered by GLAST (see Jungman, Kamionkowski and Griest, 1996,
for a general discussion of dark matter candidates).  The good energy 
resolution possible with the GLAST calorimeter will make a search for such WIMP 
annihilation lines possible.

\subsection{\underline{Pulsars}}

A number of young and middle-aged pulsars have their energy output dominated by 
their gamma-ray emission.  Because the gamma rays are directly related to the 
particles accelerated in the pulsar magnetospheres, they give specific 
information about the physics in these high magnetic and electric fields. 
Models based on the EGRET-detected pulsars make specific predictions that will 
be testable with the larger number of pulsars that GLAST's greater sensitivity 
will provide (Thompson, et al. 1997).

\subsection{\underline{Supernova Remnants and the Origin of Cosmic Rays}}

Although a near-consensus can be found among scientists that the high-energy 
charged particle cosmic rays originate in supernova remnants (SNR), the proof 
of that hypothesis has remained elusive.  Some EGRET gamma-ray sources appear 
to be associated with SNR, but the spatial and spectral resolution make the 
identifications uncertain (e.g. Esposito et al. 1996).  If SNR do accelerate 
cosmic rays, they should produce gamma rays at a level that can be studied with 
GLAST, which will be able to resolve some SNR spatially. 

\subsection{\underline{Diffuse Gamma Radiation}}

Within the Galaxy, GLAST will explore the diffuse radiation on scales from 
molecular clouds to galactic arms, measuring the product of the cosmic ray and 
gas densities.  The extragalactic diffuse radiation may be resolved; GLAST 
should detect all the blazars suspected of producing this radiation.  Any 
residual diffuse extragalactic gamma rays would have to come from some new and 
unexpected source. 

\subsection{\underline{Unidentified Sources and New Discoveries}}

Over half the sources seen by EGRET in the high-energy gamma-ray sky remain 
unidentified with known astrophysical objects.  Some may be radio-quiet 
pulsars, some unrecognized blazars, and some are likely to be completely new 
types of object (for a recent discussion, see Mukherjee et al., 1997). In 
general, the EGRET error boxes are too large for spatial correlation, and the 
photon density is too small for detailed timing studies.  Both these 
limitations will be greatly alleviated with GLAST.  In particular, the 
combination of GLAST with the next generation of X-ray telescopes should 
resolve a large part of this long-standing mystery. The new capabilities of 
GLAST will surely produce unanticipated discoveries, just as each previous 
generation of gamma-ray telescope has done.

\section{GLAST HARDWARE DEVELOPMENT}
\subsection{\underline{Glast Technologies}}
\begin{figure}
\centerline{\psfig{figure=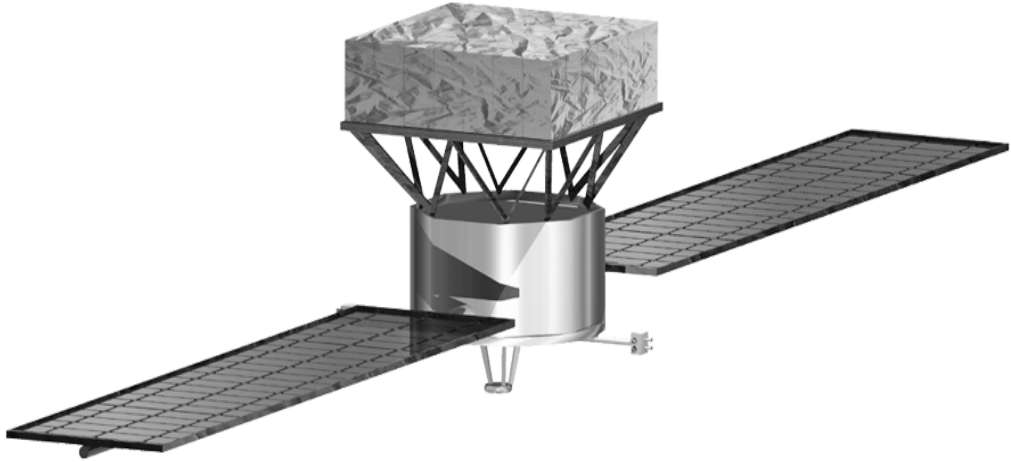,height=7cm,bbllx=50pt,bblly=150pt,bburx=750pt,bbury=500pt
,clip=.}}
\caption{A visualization of the proposed GLAST instrument and spacecraft. The 
picture shows the current baseline instrument consisting of 25 tower modules.}
\end{figure}

Any high-energy gamma-ray telescope operates in the range where pair 
production is the dominant energy loss process; therefore, GLAST (see Fig. 1 
for one concept configuration) shares some design heritage with SAS-2, COS-B, 
and EGRET: it has a plastic anticoincidence system, a tracker with thin plates 
of converter material, and an energy measurement system. What GLAST benefits 
from most is the rapid advance in semiconductor technology since the previous 
gamma-ray missions.  The silicon revolution affects GLAST in two principal ways
as will be described below.

\subsubsection{\underline{Multi-Layer Si Strip Tracker}}
The tracker consists of solid-state devices instead of a gas/wire chamber.  
The baseline design for GLAST uses Si strip detectors with 195 $\mu$m pitch
(see Fig. 2), offering significantly better track resolution with no 
expendable gas or high voltage discharge required.  Low-power application 
specific integrated circuits (ASICs) allow readout of approximately 10$^6$ 
channels of tracker with only 260 W.

The 77 m$^{2}$ of Si strip detectors planned for GLAST will be the largest Si strip detector
system ever made. Since manufacturers (Hamamatsu, Micron and others) have 
decided to move to 6-inch wafers, we can expect a good cost/performance. 

\subsubsection{\underline{On-board Computer}}
On-board computing, which was extremely limited in the Compton Observatory 
era, is now possible on a large scale.  The 32-bit, radiation-hard processors 
now available allow software to replace some of the hardware triggering of 
previous missions and also enable considerable on-board analysis of the 
tracker data to enhance the throughput of useful gamma-ray data.

\begin{figure}
\centerline{\psfig{figure=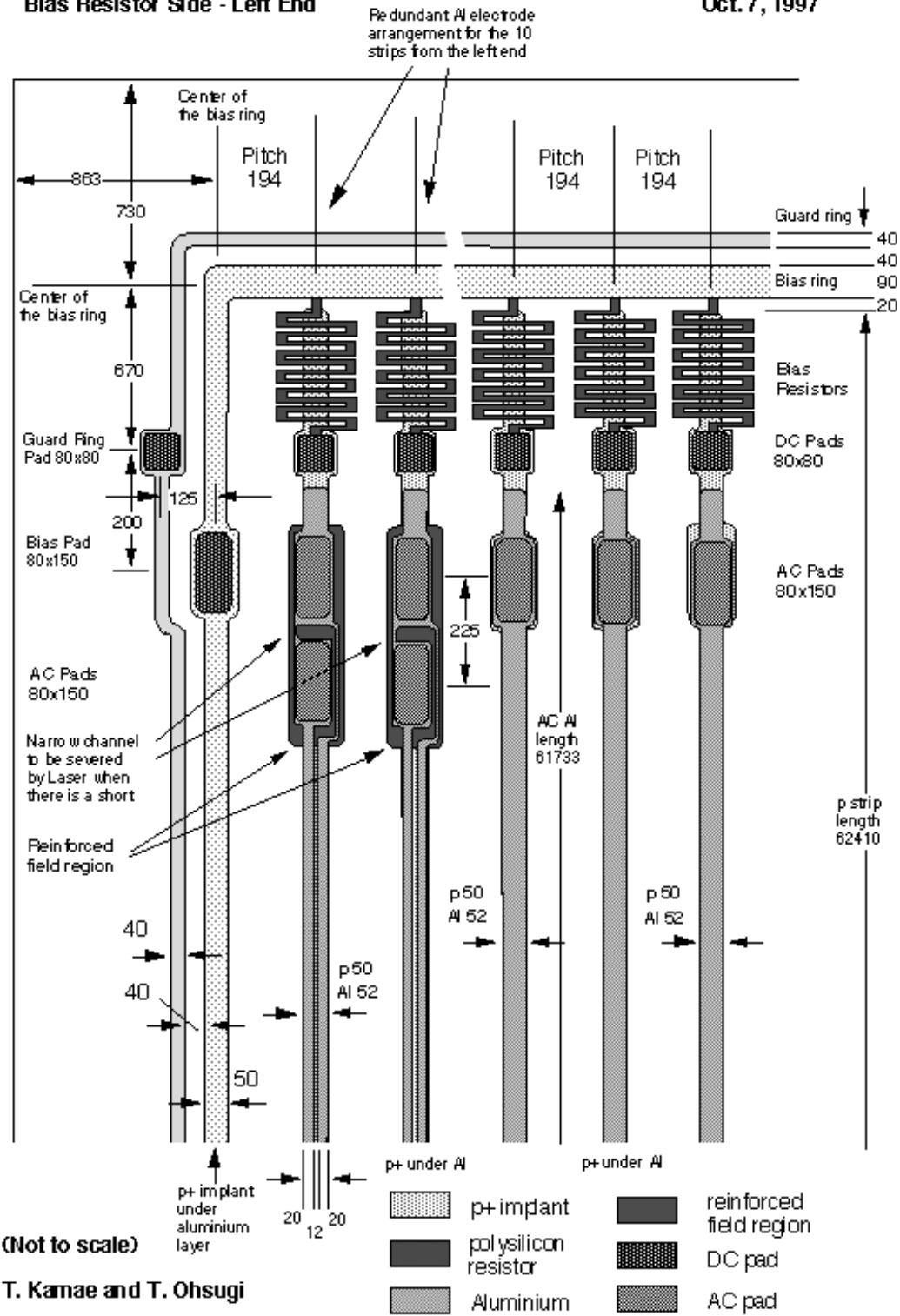,height=22cm,bbllx=00pt,bblly=00pt,bburx=550pt,bbury=800pt
,clip=.}}
\caption{Silicon Strip Detector for the GLAST Prototype Tower. The linear 
dimension is in $\mu$m.}
\end{figure}

\subsection{\underline{Plan and Schedule}}
Two prototype Si strip detectors have been made. The first prototype was a 6-cm-detector. The second prototype (Fig. 2), which had redundancy strips and 
bypass strips, showed superb characteristics (bad strips $<$ 0.03\%). A third
prototype, made from 6-inch-wafer, is being requested. 

A ``mini-tower'' consisting of a stack of the 6 cm Si strip detectors, a CsI(Tl) calorimeter, and a plastic scintillator anticoincidence, was tested at a tagged gamma-ray beam at SLAC in Fall, 1997 (Ritz, et al., 1998). A full prototype tower will be 
built by summer, 1999, for the Fall, 1999, beam test.  We are expecting full production of the GLAST hardware to begin in 2001.      
We are currently waiting for approvals from DOE and NASA. We will also apply
for the grant-in-aids from Monbusho (the Ministry of Education of Japan).

\section{THE GLAST COLLABORATION}

GLAST is planned as a facility-class mission involving an international 
collaboration of the particle physics and astrophysics communities.  Currently,
scientists from the United States, Japan, France, Germany, and Italy are 
involved in the development effort.  GLAST is currently listed as a candidate 
for a new start at NASA, with a possible launch in 2005. Further information 
about GLAST can be found at

http://www-glast.stanford.edu/

\section{REFERENCES}
\vspace{-5mm}
\begin{itemize}
\setlength{\itemindent}{-8mm}
\setlength{\itemsep}{-1mm}

\item[] 
Bloom, E.D., {\it Sp. Sci. Rev.}, {\bf 75}, 109 (1996)

\item[] 
Dingus, B.L., {\it Ap. \& Sp. Sci.}, {\bf 231}, 187 (1995)

\item[] Djorgovski, S.G. {\it et al.}, {\it IAU Circ.}, 6655 (1997)

\item[] Esposito, J.A. {\it et al.}, {\it ApJ}, {\bf 461}, 820 (1996)

\item[] Hartman, R.C., Collmar, W., von Montigny, C., \& Dermer, C.D., in 
{\it Proc. Fourth Compton Symposium}, ed. C.D. Dermer, M.S. Strickman, J.D. 
Kurfess, pp. 307--327, AIP CP410, Woodbury, NY (1997).

\item[] 
Jungman, G., Kamionkowski, M., \& Griest, K., {\it Phys. Reports}, {\bf 267}, 
195 (1996)

\item[] 
Macminn, D. \& Primack, J.R., {\it Sp. Sci. Rev.}, {\bf 75}, 413 (1996)

\item[] 
Michelson, P.F., {\it SPIE}, {\bf 2806}, 31 (1996)

\item[] Mukherjee, R., Grenier, I.A., \& Thompson, D.J., in {\it Proc. Fourth 
Compton Symposium}, ed. C.D. Dermer, M.S. Strickman, J.D. Kurfess, pp. 
394--406, AIP CP410, Woodbury, NY (1997)

\item[] Ritz, S.M. , {\it et al.}, {\it NIM}, submitted (1998)

\item[] Shrader, C.R. \& Wehrle, A.E., in {\it Proc. Fourth Compton Symposium},
ed. C.D. Dermer, M.S. Strickman, J.D. Kurfess, pp. 328--343, AIP CP410, 
Woodbury, NY (1997)

\item[] Thompson, D.J., Harding, A.K., Hermsen, W., \& Ulmer, M.P., {\it Proc. 
Fourth Compton Symposium}, ed. C.D. Dermer, M.S. Strickman, J.D. Kurfess, pp. 
39--56,  AIP CP410, Woodbury, NY (1997)

\end{itemize}

\end{document}